\documentclass[sn-basic,iicol]{sn-jnl}

\usepackage{graphicx}%
\usepackage{multirow}%
\usepackage{amsmath,amssymb,amsfonts}%
\usepackage{amsthm}%
\usepackage{mathrsfs}%
\usepackage[title]{appendix}%
\usepackage{xcolor}%
\usepackage{textcomp}%
\usepackage{manyfoot}%
\usepackage{booktabs}%
\usepackage{algorithm}%
\usepackage{algorithmicx}%
\usepackage{algpseudocode}%
\usepackage{listings}%
\usepackage{natbib}
\usepackage{multirow}%
\usepackage{multicol}
\setcitestyle{aysep={},yysep={,},notesep={:},citepunct={;}}

\theoremstyle{thmstyleone}%
\theoremstyle{thmstyletwo}%

\theoremstyle{thmstylethree}%

\begin{document}

\title[Article Title]{Constraining the emergent dark energy models with observational data at intermediate redshift}


\author[1]{\fnm{GuangZhen} \sur{Wang}}\email{guangzhenwang@gznu.edu.cn}

\author[2]{\fnm{Xiaolei} \sur{Li}}\email{lixiaolei@hebtu.edu.cn}

\author*[1,3]{\fnm{Nan} \sur{Liang}}\email{liangn@bnu.edu.cn}

\affil*[1]{\orgdiv{Key Laboratory of Information and Computing Science Guizhou Province (School of Cyber Science and Technology)}, \orgname{Guizhou Normal University}, \orgaddress{\city{Guiyang}, \postcode{Guizhou 550025}, \country{China}}}

\affil[2]{\orgdiv{College of Physics}, \orgname{Hebei Normal University}, \orgaddress{\city{Shijiazhuang}, \postcode{Heibei 050024}, \country{China}}}


\affil[3]{\orgdiv{Joint Center for FAST Sciences Guizhou Normal University Node}, \orgaddress{\city{Guiyang}, \postcode{Guizhou 550025}, \country{China}}}


\abstract{In this work, we investigate the phenomenologically emergent dark energy (PEDE) model and its generalized form, namely the generalized emergent dark energy (GEDE) model, which introduces a free parameter \unboldmath {\( \Delta \)} that can discriminate between the \unboldmath{$\mathrm{\Lambda}$}CDM model and the PEDE model.
Fitting the emergent dark energy (EDE) models with the observational datasets including the cosmology-independent gamma-ray bursts (GRBs) and the observational Hubble data (OHD) at intermediate redshift, we find a large value of $H_0$
which is close to the results of local measurement of $H_0$ from the SH0ES Collaboration in both EDE models.
In order to refine our analysis and tighten the constraints on cosmological parameters, we combine mid-redshift observations GRBs and OHD with baryon acoustic oscillations (BAOs). Finally, we constrain DE models by using the simultaneous fitting method, in which  the parameters of DE models and the relation parameters of GRBs are fitted simultaneously.
Our results suggest that PEDE and GEDE models can 
be possible alternative to the standard cosmological model, pending further theoretical explorations and observational verifications.
}




\maketitle

\section{Introduction}
One of the crucial cosmological discoveries was the late-time accelerated expansion of the universe \citep{1998AJ....116.1009R,1999ApJ...517..565P}, a phenomenon that remains mysterious within the current cosmological framework. 
To provide a plausible explanation, the concept of an exotic cosmic component dark energy (DE) which produces negative pressure  with a negative equation of state was introduced.
The late-time accelerated expansion of the universe can be modeled by the $\Lambda$CDM model,
which combining the simplest assumption for dark energy: the cosmological constant $\Lambda$ with an equation of state (EoS) parameter \(w = -1\)
and the cold dark matter (CDM) component. 
The standard $\Lambda$CDM model 
has successfully described numerous cosmological observations, including Type Ia supernovae (SNe Ia) \citep{2007ApJ...659...98R,2018ApJ...859..101S}, baryon acoustic oscillations (BAOs) \citep{2011MNRAS.416.3017B,2015MNRAS.449..835R,2017PhRvD..95b3524A,2019MNRAS.482.3497Z,2021PhRvD.103h3533A}, and the cosmic microwave background (CMB) \citep{2009PhRvD..80h3002I,2011ApJS..192...18K,2016A&A...594A..13P,2020A&A...641A...6P}.
The measurement of the Hubble constant ($H_0$) has revealed the current accelerated expansion of the Universe \citep{2010ARA&A..48..673F}.
The $H_0$ tension is one of the major issues in modern cosmology in which  the measurements discrepancy  between the local measurement of $H_0$ by the  \emph{Supernova H0 for the Equation of
State} (SH0ES) collaboration \citep{2016ApJ...826...56R,2018ApJ...853..126R,2019ApJ...876...85R,2022ApJ...934L...7R,2022ApJ...938...36R}  and the early Universe using Planck CMB 
observations assuming the $\Lambda$CDM model \citep{2016A&A...594A..13P,2020A&A...641A...6P} can reach at $5.3\sigma$. 
At a 1-$\sigma$ confidence level, SH0ES measurement of the distance ladder calibrated by Cepheids 
yields $H_0 = 73.01 \pm 0.99 \, \mathrm{km \, s^{-1} Mpc^{-1}}$ \citep{2022ApJ...938...36R};
whereas the Planck collaboration which uses temperature and polarization anisotropies in the CMB obtain $H_0 = 67.27 \pm 0.6\, \mathrm{km \, s^{-1} Mpc^{-1}}$ \citep{2020A&A...641A...6P}.
The $H_0$ tension implies that either there are considerable but not accounted for systematic errors in observations, or modifications to the standard $\Lambda$CDM model might be considered, see \citep{2007MNRAS.376.1767O,2007PhRvD..75h3517B,2021CQGra..38o3001D} and reference therein.

With a motivation of alleviating the $H_0$ tension,
\cite{Li_2019} proposed a new dark energy model called the Phenomenologically Emergent Dark Energy (PEDE) model as a potential alternate to the $\mathrm{\Lambda CDM}$ model. 
The model effectively replaces the cosmological constant with a hyperbolic tangent function of redshift which causes the DE to emerge as a function of the cosmic time at later times. 
\cite{2020JCAP...06..062P}  found that the tension on $H_0$ is clearly alleviated for the PEDE model in a six parameter space  similar to the spatially flat $\mathrm{\Lambda CDM}$  model  with the combined datasets. 
\cite{2020ApJ...899....9K} used  a nonparametric iterative smoothing method on the Joint Light-curve Analysis (JLA) SNe Ia data to show that the PEDE model are consistent with those of the standard model.
 \cite{2023PhRvD.107f3509Y} considered the effects of adding curvature in the PEDE model with the Planck 2018 CMB temperature and polarization data, BAO and Pantheon sample \citep{2018ApJ...859..101S} which contains 1048 SNe Ia data.
\cite{2022A&A...668A..51L} used  a newly compiled sample the ultra-compact structure in radio quasars and strong gravitational lensing systems with quasars to constrain the spatially flat and non-flat PEDE model.

Later on, \cite{Li_2020} proposed the Generalized Emergent Dark Energy (GEDE) model  with extra parameters to describe the properties of dark energy evolution:
the free parameter $\Delta$ describe the evolution slope of dark energy density, and the transition redshift $z_t$  which identifies where dark energy density equals matter density  is not a free parameter.
The GEDE model has the flexibility to include both the $\mathrm{\Lambda CDM}$ model and the PEDE model as two of its special limits.
\cite{2021Univ....7..163M} briefly summarize the characteristics of a list of dark energy models including the PEDE and GEDE models with the joint cosmological samples.

There is an interesting idea for the $H_0$ tension for $H_0$ with a redshift evolving of observational data.
Recently, \cite{2021ApJ...912..150D} find a slowly decreasing trend of $H_0$ value with a function mimicking the redshift evolution. The local distance ladder of SN Ia calibrated by Cepheids can reach at $z<0.01$, while the CMB data is near $z\sim1000$. Therefore, cosmological data in the mid-redshift region between the local distance ladder and CMB might offer important insights into the origins of the $H_0$ tension.
Gamma-ray bursts (GRBs) are extremely powerful and bright sources that are observed up to very high redshifts, reaching at \( z = 8.2 \) \citep{2009Natur.461.1254T} and \( z = 9.4 \) \citep{2011ApJ...736....7C}.
Therefore, GRBs can be used to probe the high-redshift universe beyond SNe Ia.
Due to the lack of a low-redshift sample, a fiducial cosmological model should be assumed  for calibrating the GRB luminosity relations in the early cosmological studies \citep{2004ApJ...612L.101D}. The so-called circularity problem  \citep{2006NJPh....8..123G} will be encountered. 
For the purpose to avoid the circularity problem, 
\cite{2008ApJ...685..354L} proposed a cosmological model-independent method to calibrate the luminosity relations of GRBs by using the SNe Ia data \citep{PhysRevD.81.083518,2011A&A...527A..11L,2010JCAP...08..020W,2016A&A...585A..68W,2022ApJ...935....7L}.

On the other side, the observational Hubble data (OHD) using the cosmic chronometers (CC) method from the galactic age differential method \citep{2002ApJ...573...37J} has advantages in constraining cosmological parameters and distinguishing DE models.
This method allows for Hubble information to be directly derived from observations up to approximately \( z \lesssim2 \) \citep{2022LRR....25....6M}.
\cite{2019MNRAS.486L..46A} proposed an alternative method to calibrate 193 GRBs
(spectral parameters taken from \cite{2017A&A...598A.112D} and references therein) with firmly measured redshift  by using the OHD with the CC method.
\cite{2023MNRAS.521.4406L} calibrated GRBs from the latest OHD using Gaussian Process to construct the GRB Hubble diagram.
\cite{2023arXiv230716467X} obtain a larger $\Omega_M$ values in the $\mathrm{\Lambda CDM}$  model with GRBs at high redshift, but adding OHD at low redshit removes this trend.
\cite{2023A&A...674A..45J} indicate that $H_0$ value 
is consistent with that measured from the local data at low redshift and drops to the value measured from the CMB at high redshift  with SN Ia, OHD and BAO data.

Recently, \cite{2020MNRAS.497.1590H} 
constrained the PEDE and GEDE models with the latest OHD, including non-homogeneous, homogeneous and differential age Hubble data,
to obtain values for the deceleration-acceleration transition redshift within a $2\sigma$ confidence level. 
More recently, \cite{2022ApJ...941...84L} used a Gaussian Process to  calibrate the A118 GRB sample from the Pantheon sample and constrained DE models with GRBs at high redshift and OHD.
In this work, we use the cosmology-independent GRBs in Ref. \citep{2022ApJ...941...84L} at $1.4<z\leq8.2$
and the latest OHD obtained with the CC method which summarized in Ref. \citep{2023MNRAS.521.4406L} at $0.07<z<1.965$ to study 
the two emergent DE models: PEDE and GEDE.
We use the information criterion DIC 
to compare the dark energy models.

The paper is organized as follows: In Section. \ref{section2}, we summarize the cosmological models to be analyzed. In Section. \ref{section3}, we briefly describe the observational data sets we used in this work and the corresponding analysis method. The results are shown in Section. \ref{section4}. Finally, the conclusions are given in Section. \ref{section5}. 

\section{Cosmological models\label{section2}}
Considering a spatially flat, homogeneous, and isotropic universe and the Friedmann-Lema\^{\i}tre-Robertson-Walker (FLRW) metric, the Friedmann equation can describe the evolution of the Universe with negligible radiation, pressureless matter, and DE:
\[
	E(a) = \left[\Omega_{\mathrm{m,0}} \times a^{-3} + \widetilde{\Omega}_{\mathrm{DE}}(a)\right]^{-\frac{1}{2}}, \tag{1}
 \]
where the scale factor $a = 1/(1 + z)$, $\Omega_{\mathrm{m,0}}$ is the current density of matter at redshift $z = 0$. $\widetilde{\Omega}_{\mathrm{DE}}(a)$ is the energy density of the dark energy fluid with respect to the critical energy density at present, with $\rho_{\mathrm{crit,0}} = 3H_0^2 / 8\pi G$ and $\rho_{\mathrm{crit}}(a) = 3H^2(a) / 8\pi G$.
The present values of the density parameters for 
pressureless matter are defined as 
$\Omega_{\mathrm{m,0}} = \rho_{\mathrm{m,0}} / \rho_{\mathrm{crit,0}}$. $\widetilde{\Omega}_{\mathrm{DE}}$ is the density of dark energy,
which is defined as:
\begin{align}
	\widetilde{\Omega}_{\mathrm{DE}}(a) &= \frac{\rho_{\mathrm{DE}}(a)}{\rho_{\mathrm{crit,0}}} \nonumber\\
	& = \frac{\rho_{\mathrm{DE}}(a)}{\rho_{\mathrm{crit}}(a)} \times \frac{\rho_{\mathrm{crit}}(a)}{\rho_{\mathrm{crit,0}}} = \Omega_{\mathrm{DE,0}}(a) \times \frac{H^2(a)}{H_0^2}, \nonumber \tag{2}
\end{align}
where $\Omega_{\mathrm{DE,0}}$ is the current density of DE at redshift $z = 0$.
Alternatively, this equation can be expressed as a function of redshift $z$:
\[
	\widetilde{\Omega}_{\mathrm{DE}}(z) = \Omega_{\mathrm{DE,0}} \times \exp\left\{\int_{0}^{z} \frac{1 + w(z')}{1 + z'} \mathrm{d}z'\right\}, \tag{3}
 \]

The PEDE model \citep{Li_2019} has been 
proposed as a potential alternative to the $\Lambda$CDM model without additional degrees of freedom. The DE density at redshift $z$ is given by:
\[
	\widetilde{\Omega}_{\mathrm{DE}}(z) = \Omega_{\mathrm{DE,0}} \times \left[1 - \tanh\left(\log_{10}(1+z)\right)\right]. \tag{4}
 \]

By assuming a more generalized form of EDE model including extra parameters,  the DE density in the GEDE model \citep{Li_2020} is given by:
\[
	\widetilde{\Omega}_{\mathrm{DE}}(z) = \Omega_{\mathrm{DE,0}} \times \frac{1 - \tanh\left(\Delta \times \log_{10}\left(\frac{1+z}{1+z_t}\right)\right)}{1 + \tanh\left(\Delta \times \log_{10}(1+z_t)\right)}, \tag{5}
 \]
where 
$z_t$ is the transition redshift, which can be derived from $\widetilde{\Omega}_{\mathrm{DE}}(z_t) = \Omega_{\mathrm{m,0}}(1 + z_t)^3$.
In the GEDE model, setting $\Delta = 0$ recovers the $\Lambda$CDM model, while setting $\Delta = 1$ yields the PEDE model, with the exception that the authors \citep{Li_2020} set $z_t = 0$ for simplicity.

In this work, we also consider the $\Lambda$CDM model, 
the $w$CDM model 
and the Chevallier-Polarski-Linder (CPL) parameterization to consider a DE component that depends on redshift \citep{2001IJMPD..10..213C,2003PhRvD..68h3503L,2005PhRvD..72d3529L,2006PhLB..635...61B} for comparison. 
The EoS of all the DE models
can be summerized as follows:
\[
\tiny
w(z) = 
\begin{cases}
	-1, & \text{$\Lambda$CDM} \\
	w_0, & \text{$w$CDM} \\
	w_0 + \frac{w_a z}{1+z}, & \text{CPL} \\
	-\frac{1}{3\ln 10} \times \left(1 + \tanh\left[\log_{10}(1+z)\right]\right) - 1, & \text{PEDE} \\
	-\frac{\Delta}{3\ln 10} \times \left(1 + \tanh\left[\Delta \times \log_{10}\left(\frac{1+z}{1+z_t}\right)\right]\right) - 1, & \text{GEDE}
\end{cases} \tag{6}
\]

In order to facilitate model comparison and evaluate their relative merits, several well-established statistical measures were employed. These included the Akaike Information Criterion (AIC) \citep{1100705}, the Bayesian Information Criterion (BIC) \citep{schwarz1978estimating}, and the Deviance Information Criterion (DIC) \citep{kunz2006measuring} - all of which have found widespread application in astrophysical research.
Since the AIC and BIC criteria employ only the likelihood value at maximum 
numerically from the Bayesian analysis, one needs to use sufficiently long chains to ensure the accuracy of $\mathcal{L}_{max}$. 
The quantity DIC, known also as the Bayesian complexity, which focus on assessing the number of parameters that can be constrained by a particular dataset, has been introduced into astrophysics.
The use of DIC can provide all the information obtained from the likelihood calls during the maximization procedure.
For a quantitative comparison between our proposed  in this work, we employ the DIC which is defined as \citep{spiegelhalter64van}:
\[
\mathrm{DIC}=D(\bar{\theta}) + 2p_D = \overline{D(\theta)}+ p_D \tag{7}
\]
where $D(\theta) = -2ln \mathcal{L}(\theta)+C$, $C$ is a normalized constant depending only on the data which will vanish from any derived quantity, $p_D = \overline{D(\theta)} - D(\bar{\theta})$ is the effective number of model parameters, with the deviance of the likelihood.


\section{Observational data\label{section3}} 
\begin{figure*}[htbp]
	\centering
	\includegraphics[width=0.45\textwidth]{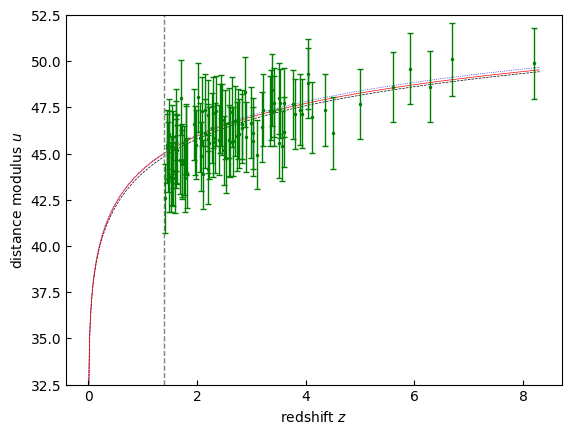}
	\includegraphics[width=0.45\textwidth]{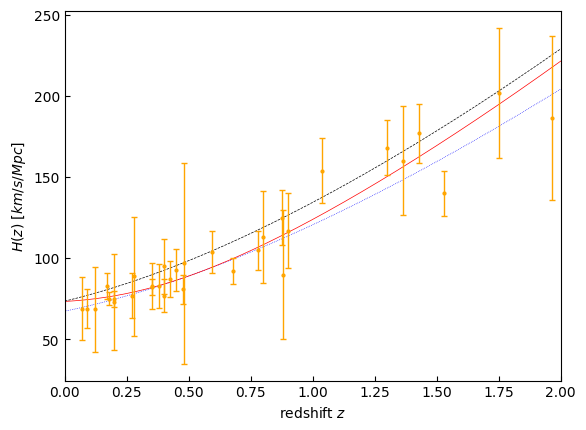}
	\caption{The  cosmology-independent 98 GRBs at $1.4<z\leq8.2$ (\emph{left}) and 32 OHD at $0.07 < z < 1.965$ (\emph{rihgt}).
The red solid curve present the predicted values from the best values of GEDE model with GRBs and OHD. The blue dotted curve and the black dashed curve are the predicted values of  distance modulus for a flat $\Lambda$CDM model from CMB and SNe Ia, respectively.\label{fig-Hubble-diagram}}
\end{figure*}

In this section, we describe the observational data used in our analyses for constraining cosmological parameters. 
For the  GRBs sample,
we follow the cosmology-independent approach in \cite{2022ApJ...941...84L} to calibrate the Amati relation with the A118 GRB sample  \citep{2021JCAP...09..042K} using the Pantheon SNe Ia sample \citep{2018ApJ...859..101S}; 
and use GRBs data at redshifts  $1.4<z\leq8.2$ to constrain cosmological models \footnote{It should be noted that  the calibration results can be affected by the treatment of absolute magnitude M. We find that the calibration parameters with the A118 GRB data set are almost the same with and without marginalization over M.}.
OHD obtained using the CC method relates the evolution of differential ages of passive galaxies at different redshifts without assuming any cosmological model \citep{2002ApJ...573...37J}. 
We utilize 32 updated OHD measurements compiled from Ref. \citep{2023MNRAS.521.4406L}, covering a redshift range of $0.07 < z < 1.965$, 
which consists of 15 correlated measurements with the corresponding covariance matrix provided by  \cite{2020ApJ...898...82M}, and 17 uncorrelated measurements with the latter sources from \citep{2012JCAP...08..006M,2015MNRAS.450L..16M,2016JCAP...05..014M}.
The  cosmology-independent 98 GRBs at $1.4<z\leq8.2$ and 32 OHD at $0.07 < z < 1.965$  are showed in Fig. \ref{fig-Hubble-diagram}.

\unskip

The cosmological parameters are fitted with GRBs by minimizing the $\chi^2$ method:
\[ \chi_{\mathrm{GRB}}^2 =  \sum_{i=1}^{N} \left[ \frac{\mu_{\mathrm{obs}}(z_i)- \mu_{\mathrm{th}}(z_i;p,H_0)}{\sigma_{\mu_i}} \right]^2 \tag{8}
\]
where $N=98$ represents the number of GRBs at high-redshift, $\mu_{\mathrm{obs}}$ is the observed distance modulus and
$\sigma_{\mu_i}$ is the uncertainty associated with the observed distance modulus,
$\mu_{\mathrm{th}}$ denotes the theoretical distance modulus 
which determined by the cosmological parameters $p$ with DE models and $H_0$.
To constrain the dark energy models using OHD, the corresponding \( \chi^2_{\mathrm{OHD}} \) is given by: \citep{2024MNRAS.530.4493Z}
\[
\chi_{\mathrm{OHD}}^2 = \sum_{i=1}^{17} \left[\frac{H_{\mathrm{obs}}(z_i) - H_{\mathrm{th}}(z_i;p,H_0)}{\sigma_{H_i}} \right]^2 + \Delta \hat{H}^{T}C_{H}^{-1} \Delta \hat{H} \tag{9}
\]
where 
$\sigma_{H_i}$ represents the observed uncertainty of the 17 uncorrelated measurements, $\Delta \hat{H}=H_{\mathrm{obs}}(z) - H_{\mathrm{th}}(z;p,H_0)$ represents the difference vector between the observed data and the theoretical values for the 15 correlated measurements, and $C_{H}^{-1}$ is the inverse of the covariance matrix.
The combine $\chi^2$ statistic, combining GRBs and OHD is given by:
\[
\chi_{\mathrm{com}}^2 = \chi_{\mathrm{GRB}}^2 + \chi_{\mathrm{OHD}}^2  \tag{10}
\]

\section{Results\label{section4}}  
\subsection{Results from GRBs and OHD}

In this section, we estimate and compare the parameters of the standard $\Lambda$CDM model, the $w$CDM model, the CPL, the PEDE model and the GEDE model using cosmological observation data from GRBs and OHD.
Through the minimization of the $\chi^2$ value, we can obtain the best-fit parameter estimates. We employ the \emph{emcee} Python module \citep{2013PASP..125..306F} in the lmfit python library \citep{matt_newville_2021_4516644}. Furthermore, we utilize the GETDIST package \citep{2019arXiv191013970L} to analyze the sampled chains.

\begin{table*}[htbp]
	\caption{Constraints at $68 \%$ confidence-level errors on the cosmological parameters for the different tested dark energy models with GRBs-only, OHD-only and GRBs + OHD. And at $95 \%$ confidence-level errors on the $\Delta$ for GEDE. }
	\tiny
	\setlength{\tabcolsep}{1pt}
	\begin{tabular*}{\textwidth}{@{\extracolsep\fill}lcccccccc@{\extracolsep\fill}}
		\toprule
		\toprule%
		Parameters & $H_0$ & $\Omega_m$ & $w_0$ & $w_a$ & $\Delta$ & $z_t^*$ & $\chi_{md}^2$ & $\mathrm{\Delta DIC}$ \\
		\midrule
		GRBs-only\\
		\midrule
		$\mathrm{\Lambda CDM}$ & $72.0_{-20.0}^{+10.0}$ & $0.50_{-0.36}^{+0.18}$ & - & - & 0 & $0.060_{-0.500}^{+0.370}$ & $26.831$ & $0$ \\[1ex]
		$w\mathrm{CDM}$ & $70.0_{-20.0}^{+8.0}$ & $0.50_{-0.36}^{+0.24}$ & $-0.98\pm0.55$ & - & - & $-0.120_{-0.490}^{+0.570}$ & $26.951$ & $+0.368$ \\[1ex]
		$\mathrm{CPL}$ & $69.0_{-20.0}^{+9.0}$ & $0.53_{-0.34}^{+0.39}$ & $-1.03_{-0.83}^{+0.42}$ & $-0.20_{-2.30}^{+1.30}$ & - & $-0.110_{-0.360}^{+0.580}$ & $26.954$ & $+0.625$ \\[1ex]
		$\mathrm{PEDE}$ & $73.0_{-20.0}^{+10.0}$ & $0.52_{-0.36}^{+0.22}$ & - & - & 1 & $0.020\pm0.360$ & $26.884$ & $-0.195$ \\[1ex]
		$\mathrm{GEDE}$ & $73.0_{-20.0}^{+10.0}$ & $0.55_{-0.31}^{+0.20}$ & - & - & $4.9\pm2.9(_{-4.6}^{+4.8})$ & $-0.001_{-0.250}^{+0.220}$ & $26.942$ & $+0.204$ \\[1ex]
		\midrule
		OHD-only\\
		\midrule
		$\mathrm{\Lambda CDM}$ & $68.8\pm4.1$ & $0.324_{-0.074}^{+0.048}$ & - & - & 0 & $0.290\pm0.120$ & $14.526$ & $0$ \\[1ex]
		$w\mathrm{CDM}$ & $70.2_{-6.7}^{+5.6}$ & $0.294_{-0.060}^{+0.084}$ & $-1.15_{-0.57}^{+0.46}$ & - & - & $0.220_{-0.140}^{+0.180}$ & $15.080$ & $+0.043$ \\[1ex]
		$\mathrm{CPL}$ & $70.5_{-6.8}^{+5.7}$ & $0.305_{-0.072}^{+0.100}$ & $-1.17_{-0.66}^{+0.40}$ & $-0.30_{-2.20}^{+1.30}$ & - & $0.270_{-0.200}^{+0.140}$ & $15.160$ & $+0.322$ \\[1ex]
		$\mathrm{PEDE}$ & $69.9\pm4.2$ & $0.332_{-0.068}^{+0.046}$ & - & - & 1 & $0.235\pm0.099$ & $14.497$ & $-0.064$ \\[1ex]
		$\mathrm{GEDE}$ & $72.4\pm4.8$ & $0.334_{-0.063}^{+0.038}$ & - & - & $3.7_{-3.5}^{+1.4}(_{-3.8}^{+5.4})$ & $0.185_{-0.092}^{+0.062}$ & $14.752$ & $+0.589$ \\[1ex]
		\midrule
		GRBs + OHD\\
		\midrule
		$\Lambda\mathrm{CDM}$ & $69.9\pm4.0$ & $0.325_{-0.070}^{+0.049}$ & - & - & 0 & $0.290\pm0.120$ & $43.250$ & $0$ \\[1ex]
		$w\mathrm{CDM}$ & $71.2_{-6.2}^{+5.2}$ & $0.298_{-0.057}^{+0.081}$ & $-1.14_{-0.43}^{+0.53}$ & - & - & $0.220_{-0.140}^{+0.170}$ & $43.682$ & $+0.034$ \\[1ex]
		$\mathrm{CPL}$ & $71.9\pm6.1$ & $0.311_{-0.067}^{+0.092}$ & $-1.18_{-0.67}^{+0.37}$ & $-0.40_{-2.30}^{+1.20}$ & - & $0.265_{-0.190}^{+0.094}$ & $43.609$ & $+0.409$ \\[1ex]
		$\mathrm{PEDE}$ & $71.0\pm 4.1$ & $0.335_{-0.066}^{+0.045}$ & - & - & 1 & $0.231\pm0.095$ & $43.221$ & $-0.190$ \\[1ex]
		$\mathrm{GEDE}$ & $73.4\pm4.7$ & $0.335_{-0.057}^{+0.040}$ & - & - & $3.6_{-3.4}^{+1.3}(_{-3.7}^{+5.1})$ & $0.184_{-0.089}^{+0.059}$ & $43.499$ & $+0.606$ \\[1ex]
		\botrule
	\end{tabular*}
	\footnotetext{\textbf{Note}: The last column of the table display the $\Delta$DIC values relative to the $\Lambda$CDM model, derived from the same data combinations. $\chi_{md}^2$ represents the median value of $\chi^2$. The parameter $z_t^*$ is not a free parameter.}
	\label{tab-Alldatasets}
\end{table*}

The results of cosmological parameters with 1$\sigma$ uncertainties constraint with GRBs-only, OHD-only and GRBs + OHD for five DE models are provided in Table \ref{tab-Alldatasets}. 
For the case with GRBs-only, 
we obtain $H_0$ and 
$\Omega_{m}$ with the large error bars which indicate that the cosmological parameters are not well-constrained with this datasets; 
the $\Lambda$CDM model ($w_0 = -1$,  $w_a=0$) are consistent with the inferred value of $w_0=-0.98\pm0.55$ for the $w$CDM model and $w_0=-1.03_{-0.83}^{+0.42}$, $w_a=-0.20_{-2.30}^{+1.30}$ for the CPL model within 1$\sigma$ uncertainty.
For the case with OHD-only, we find that the value of $H_0$ for the PEDE model ($H_0=69.9\pm4.2 \, \mathrm{km \, s^{-1} Mpc^{-1}}$) is lower than that of the GEDE model ($H_0=72.4\pm4.8 \, \mathrm{km \, s^{-1} Mpc^{-1}}$), 
which shows agreement with the SH0ES measurement \citep{2022ApJ...938...36R,2022ApJ...934L...7R}.
For the case with GRBs + OHD, the measured $H_0$ ranges from $69.9 \pm 4.0 \, \mathrm{km \, s^{-1} Mpc^{-1}}$ ($\Lambda$CDM) to $73.4 \pm 4.7 \, \mathrm{km \, s^{-1} Mpc^{-1}}$ (GEDE).
When the OHD is combined with GRBs, 
we find the constraints results on $H_0$ and $\Omega_m$ can be significantly improved
and the mean values shifts in the same direction, though the overall effect is not very large.
From Table \ref{tab-Alldatasets}, we can see that for all models, the constraints on $H_0$ and $\Omega_m$ from OHD and GRBs + OHD are well consistent with each other at $1\sigma$ CL, but in agreement with the constraint from GRBs at about $2\sigma$.
Interestingly, the constraints for the $w$CDM model and CPL model are not well-constrained and exhibit results distinct from the other models. 


The statistical measures of the model comparison for the three datasets are also presented in Table \ref{tab-Alldatasets}. 
The PEDE model outperforms the $\Lambda$CDM model in both the GRBs-only and OHD-only datasets, with 
$\Delta\mathrm{DIC}=-0.195$ and $\Delta\mathrm{DIC}=-0.064$, respectively.
This trend continues in the combined GRBs + OHD dataset, where the PEDE model surpasses not only the $\Lambda$CDM model but also the $w$CDM and CPL models across all evaluation measures.
However, the GEDE model does not exhibit clear evidence of superiority over the $\Lambda$CDM, $w$CDM and CPL models in any of the datasets. It is noteworthy that this analysis is conducted without assuming any hard-cut prior on the Hubble constant ($H_0$), ensuring an unbiased comparison of the models.
In summary, the PEDE model consistently demonstrates a better fit to the data compared to the $\Lambda$CDM model, as evidenced by its lower DIC values. In contrast, the $w$CDM and CPL parameterization models perform poorly in terms of DIC when compared to the $\Lambda$CDM model, highlighting the PEDE model's superiority in describing the observations across all three datasets.

\begin{figure*}[htbp]
	\centering
	\includegraphics[width=0.33\textwidth,height=0.33\textwidth]{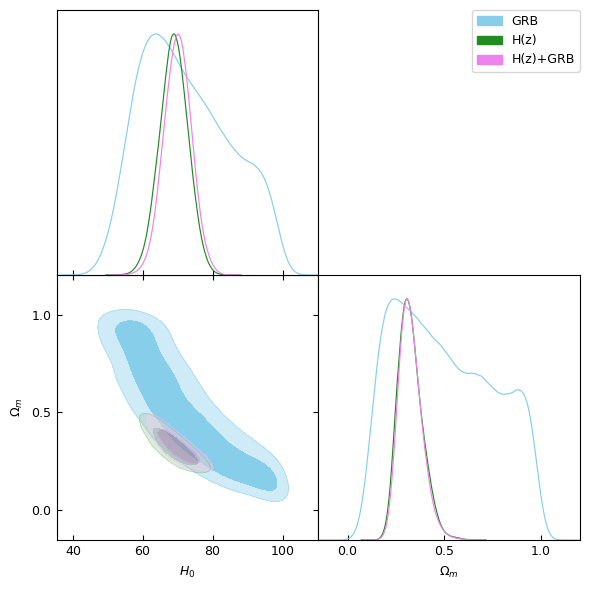}
	\includegraphics[width=0.49\textwidth]{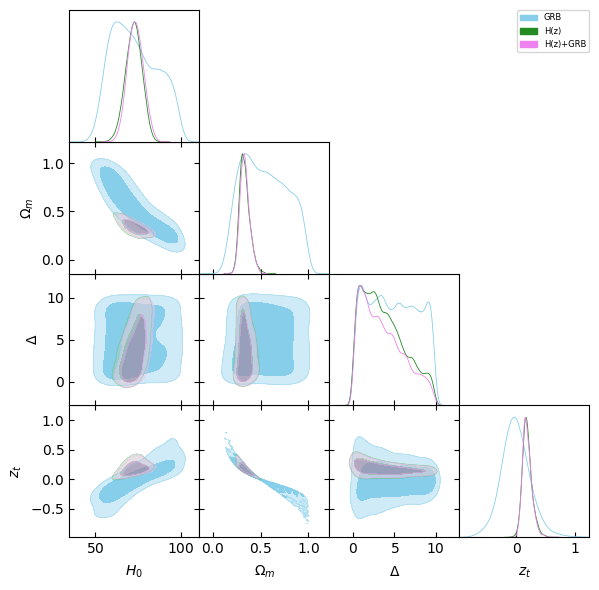}
	\caption{
Contours with $1 \sigma$ and $2 \sigma$ CL of cosmological parameters ($H_0$ and $\Omega_m$) for the $\Lambda$CDM model (\emph{left}) and cosmological parameters ($H_0$, $\Omega_m$ and $\Delta$) in the framework of GEDE (\emph{right}) from GRBs-only, OHD-only, GRBs + OHD. Note that $z_t$ is not a free parameter and is shown for clarity. \label{fig-GEDE-LCDM-ALL}}
\end{figure*}
\unskip


In Fig. \ref{fig-GEDE-LCDM-ALL},  we show the constrained results of the cosmological parameters for the $\Lambda$CDM and GEDE model with GRBs-only, OHD-only and GRBs + OHD datasets. 
We find the constraints on $H_0$ are all in agreement with each other at $1\sigma$ confidence level and also agree with the local results from the SH0ES collaboration \citep{2019ApJ...876...85R}.
For the free parameter of the GEDE model, 
we can find that the results with GEDE exclude PEDE and $\Lambda$CDM in $1\sigma$ and with large error for GRBs-only case.
$\Delta = 0$ is in agreement at about $1.7\sigma$, and $\Delta=1$ is at about $1.3\sigma$.
For OHD-only and GRBs + OHD datasets, we get the result with tight error bars and that PEDE is preferred, namely, $\Delta$ close to 1. We get $\Delta = 1$ is in agreement at about $1.9\sigma$, $\Delta=0$ is at about $2.6\sigma$ for OHD-only and $\Delta = 1$ is in agreement at about $2\sigma$, $\Delta=0$ is at about $2.8\sigma$ for GRBs + OHD. $\Lambda$CDM are excluded in $2\sigma$.
Interestingly, the derived parameter $z_t$ of the GEDE model ($z_t=0.185_{-0.092}^{+0.062}$) with OHD-only and ($z_t=0.184_{-0.089}^{+0.059}$) with  GRBs + OHD  are in agreement with the result of Hern{\'a}ndez-Almada et al. ($z_t=0.174_{-0.064}^{+0.083}$) from OHD sample \citep{2020MNRAS.497.1590H}.
Our result are aslo consist with \cite{2022A&A...668A..51L}. 

\begin{figure*}[htbp]
	\centering
	\includegraphics[width=0.33\textwidth]{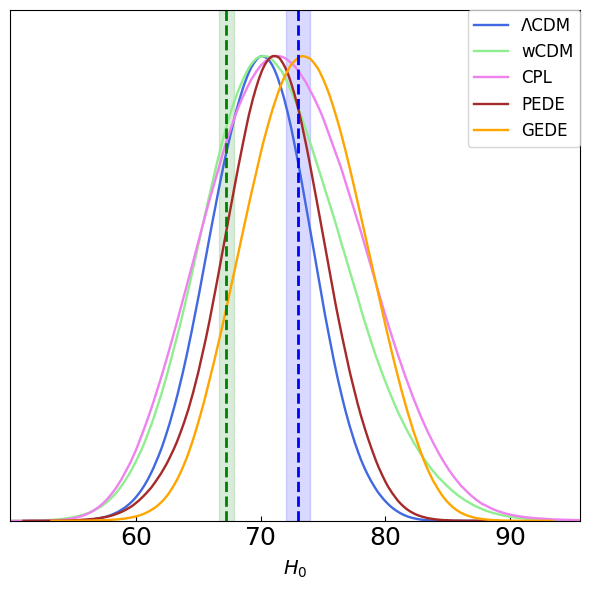}
	\includegraphics[width=0.33\textwidth]{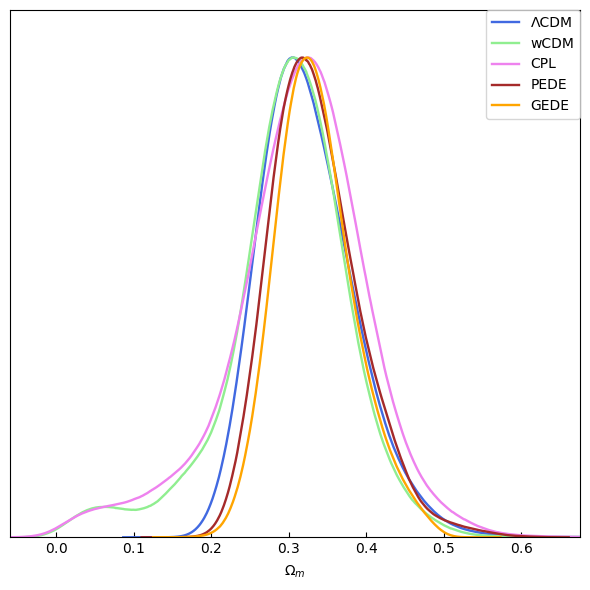}
	\caption{The constraints for $H_0$ (\emph{left}) and $\Omega_{m}$ (\emph{right}) with the $\mathrm{\Lambda CDM}$, $w$CDM, CPL, PEDE and GEDE model from the GRBs + OHD data. 
The blue line, green line and violet line represent the result from $\mathrm{\Lambda CDM}$ , $w$CDM and CPL. The brown and orange lines represent the result from PEDE and GEDE respectively. The blue shadows show the $H_0$ results with $1\sigma$ uncertainty from Riess et al. \citep{2019ApJ...876...85R}, the green shadows show the $H_0$ results with $1\sigma$ uncertainty  from Planck CMB observations \citep{2020A&A...641A...6P}. \label{fig-ev-OHD+GRB}}
\end{figure*}

In Fig. \ref{fig-ev-OHD+GRB}, we present the constraints on $H_0$ and $\Omega_{m}$ for the $\Lambda$CDM, $w$CDM, CPL, PEDE and GEDE models using the combined OHD and GRBs. 
We can find that the PEDE and GEDE models yield higher values  with a clear trend for both parameters of $H_0$ and $\Omega_{m}$ compared to $\Lambda$CDM. Furthermore, the GEDE model exhibits even more higher values than the PEDE model. These findings suggest that the EDE models have the potential to alleviate the $H_0$ tension.
It is evident that the PEDE model and GEDE model can yield a higher best-fit value of $H_0$ than the $\Lambda$CDM model when considering the GRBs-only, OHD-only and GRBs + OHD cases. These results are more consistent with those from the SH0ES collaboration \citep{2019ApJ...876...85R}.

\unskip

\begin{figure}[htbp]
	\centering
	\includegraphics[width=0.45\textwidth]{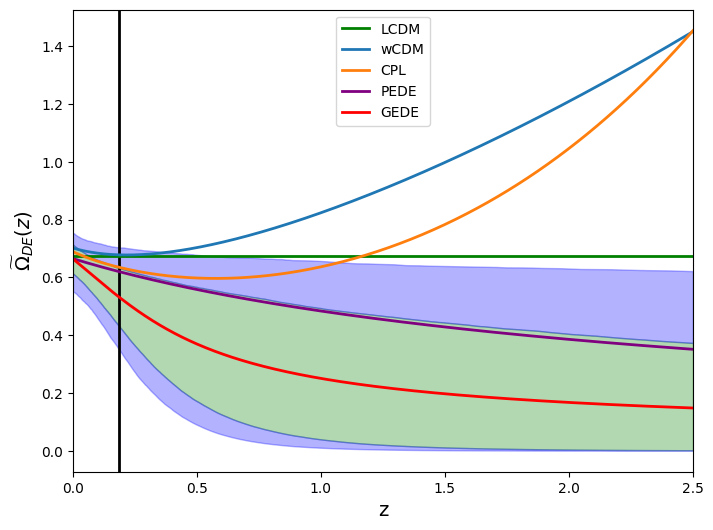}
	\caption{The evolution of dark energy density $\widetilde{\Omega}_{\mathrm{DE}}(z)$ from $z=0$ to $2.5$. 
The green and dark blue regions are the $1\sigma$ and $2\sigma$ confidence ranges of the GEDE model fitting GRBs + OHD data.
The green, blue, orange, purple and red solid lines are the best-fit results from the $\Lambda$CDM, $w$CDM, CPL, PEDE and GEDE models, respectively. 
The vertical lines display the mean values of $z_t$ from GRBs + OHD.  \label{fig-DE-evo}}
\end{figure}

In Fig. \ref{fig-DE-evo}, we show the evolution of dark energy density $\widetilde{\Omega}_{\mathrm{DE}}(z)$ as a function of redshift $z$. 
We can see an emergent dark energy behavior from GRBs + OHD data, and the cosmological constant is outside the $2\sigma$ confidence limits.

\unskip


We find our results are compatible with the previous works of Li et al. \citep{Li_2019,Li_2020}, where the authors observed that the value of $H_0$ derived from the $\Lambda$CDM and CPL parameterization models is close to the CMB prediction, regardless of whether the dataset includes CMB data or not. The authors also found that the value of $H_0$ aligns closely with the local measurement value obtained by the SH0ES collaboration when assuming $1 \sigma$ and $2 \sigma$ priors for $H_0$ taken from the SH0ES result. Our result is compatible with their findings, but it is important to emphasize that we perform our analysis without assuming any hard-cut prior on $H_0$.

\subsection{Results from GRBs, OHD and BAOs}
BAOs serve as a universal standard ruler evolving with the Universe, offering a distinct perspective on the universe's structure and evolution, which can be used as an invaluable tool for probing cosmological models \citep{2020EPJC...80..374R,2023JCAP...06..038S,2022MNRAS.509.2593R,2022A&A...668A.135S}.
In order to refine our analysis and tighten the constraints on cosmological parameters, we combine mid-redshift observations GRBs and OHD with BAOs. It should be noted that BAO measurements which under a fiducial cosmology could provide biased constraints \citep{2020MNRAS.497.1590H}. 
Here we use the 6dF Galaxy Survey (6dFGS) at $z_{\rm eff}=0.106$ \citep{2011MNRAS.416.3017B}, the Sloan Digital Sky Survey (SDSS) DR7 Main Galaxy Sample (MGS) at $z_{\rm eff}=0.15$ \citep{2015MNRAS.449..835R}, and nine measurements from the extended Baryon Oscillation Spectroscopic Survey (eBOSS) DR16 at $z_{\rm eff}=0.38,0.51,0.70,0.85,1.48$ \citep{2021PhRvD.103h3533A}. The likelihood of BAO for different datasets can be expressed as,
\[
\chi_{\mathrm{BAO}}^2 = \Delta P_{\mathrm{BAO}} C_{\mathrm{BAO}}^{-1} \Delta P_{\mathrm{BAO}}^{\mathrm{T}}, \tag{11}
\]
where $C_{\mathrm{BAO}}$ is the covariance matrix\footnote{For uncorrelated points the covariance matrix is a diagonal matrix, and its elements are the inverse errors, and for correlated points, the covariance matrices are from \citep{2021PhRvD.103h3533A}.}, $\Delta P_{\mathrm{BAO}}=v_{\mathrm{obs}}(z) - v_{\mathrm{th}}(z)$, $v_{\mathrm{obs}}(z)$ is a BAO measurement of the observed points at each $z$, and $v_{\mathrm{th}}(z)$ is the prediction of the theoretical model. The BAO feature appears in both the line-of-sight direction and the transverse direction and provides  measurements of the radial projection $D_{\mathrm{H}}$ and the transverse comoving distance $D_{\mathrm{M}}(z)$ with 
$\frac{D_{\mathrm{H}}(z)}{r_{\mathrm{d}}}=\frac{c}{H(z)r_{\mathrm{d}}}$ and
$\frac{D_{\mathrm{M}}(z)}{r_{\mathrm{d}}}=\frac{c}{H_0 r_{\mathrm{d}}}\Gamma(z)$,
where $r_{\mathrm{d}}$ is the sound horizon at the drag epoch
$r_{\mathrm{s}}(z_{\mathrm{d}})$\footnote{The comoving sound horizon \( r_{\mathrm{s}}(z) \) is given as \citep{1998ApJ...496..605E}:
$r_{\mathrm{s}} (z)=\frac{c}{H_0} \int_{z}^{\infty} \frac{c_{\mathrm{s}}}{E(z^{'}) dz^{'}}$. The redshift of the drag epoch can be approximated as \citep{1996ApJ...471..542H}:
$z_ {\mathrm{d}} = \frac{1345 \omega_m^{0.251}}{1+ 0.659 \omega_m^{0.828}}[1+b_1 \omega_b^{b_2}]$,
where $b_1=0.313 \omega_m^{-0.419}[1+0.607 \omega_m^{0.674}],b_2=0.238 \omega_m^{0.223}$ with $\omega_b=\Omega_{\mathrm{b}} h^2$ and $\omega_m=\Omega_{\mathrm{m}} h^2$.},  and $\Gamma(z)=\int_{0}^{z}dz^{'}/E(z^{'})$.
The angular diameter distance $D_{\mathrm{A}}(z)$ has relation with $D_{\mathrm{M}} (z)$: $D_{\mathrm{A}}(z)=D_{\mathrm{M}}(z)/(1+z)$.
The total $\chi^2$ statistic, combining GRBs, OHD and BAOs, is given by:
\[
\chi_{\mathrm{tot}}^2 = \chi_{\mathrm{GRB}}^2 + \chi_{\mathrm{OHD}}^2 + \chi_{\mathrm{BAO}}^2. \tag{12}
\]

Results from GRBs, OHD and BAOs for DE models are summerizd in Table \ref{tab-GRB-OHD-BAO}.
It should be noted that Rezaei et al. applied the statistical Bayesian evidence with the combining observational datasets by to indicate that the PEDE models are not favored \citep{2020EPJC...80..374R};
\cite{2023JCAP...06..038S} used the latest datasets of SNIa, CMB, and BAOs to conclude that the PEDE model cannot resolve the tension with the SH0ES measurement within $1\sigma$.
We find a significant improvement in the precision of cosmological parameter estimations, evidenced by a marked decrease in the width of the error bars, when adding BAO data into the joint analysis.
Our results  indicate that PEDE model is still a favorite model from GRBs, OHD and BAOs.
The statistical limit in our results  from GRBs, OHD and BAOs can alleviated the tension with the SH0ES measurement by $0.81 \sigma$. We obtain $H_0$ of $67.8\pm2.5 \, \mathrm{km \, s^{-1} Mpc^{-1}}$, and $\Omega_{m}=0.358_{-0.040}^{+0.034}$ for the $\Lambda$CDM model, and the two EDE models perform well our analysis with GRBs, OHD, and BAOs. These results are in consistent with those in \citep{2022A&A...668A.135S} for the $\Lambda$CDM model and  EDE models  by adding SN Ia dataset and two BAO datasets.

\begin{table*}[htbp]
	\caption{Constraints at $68 \%$ confidence-level errors on the cosmological parameters for  DE models with GRBs + OHD + BAOs. And at $95 \%$ confidence-level errors on the $\Delta$ for GEDE. }
	\tiny
	\setlength{\tabcolsep}{1pt}
	\begin{tabular*}{\textwidth}{@{\extracolsep\fill}lcccccccc@{\extracolsep\fill}}
		\toprule
		\toprule%
		Parameters & $H_0$ & $\Omega_m$ & $w_0$ & $w_a$ & $\Delta$ & $z_t^*$ & $\chi_{md}^2$ & $\mathrm{\Delta DIC}$ \\
		\midrule
		GRBs + OHD + BAO\\
		\midrule
		$\Lambda\mathrm{CDM}$ & $67.8\pm2.5$ & $0.358_{-0.040}^{+0.034}$ & - & - & 0 & $0.219\pm0.067$ & $45.070$ & $0$ \\[1ex]
		$w\mathrm{CDM}$ & $68.8_{-4.3}^{+3.6}$ & $0.358_{-0.049}^{+0.040}$ & $-1.13_{-0.17}^{+0.30}$ & - & - & $0.203_{-0.100}^{+0.073}$ & $45.190$ & $+2.275$ \\[1ex]
		$\mathrm{CPL}$ & $67.6_{-4.3}^{+3.8}$ & $0.375\pm0.052$ & $-0.85\pm0.34$ & $-0.94_{-1.70}^{+0.93}$ & - & $0.203\pm0.079$ & $44.953$ & $+2.183$ \\[1ex]
		$\mathrm{PEDE}$ & $70.6\pm2.8$ & $0.346_{-0.041}^{+0.034}$ & - & - & 1 & $0.208\pm0.059$ & $45.494$ & $+0.489$ \\[1ex]
		$\mathrm{GEDE}$ & $70.4_{-3.4}^{+3.0}$ & $0.345_{-0.041}^{+0.034}$ & - & - & $1.02_{-1.00}^{+0.41}(_{-1.00}^{+1.60})$ & $0.210\pm0.061$ & $45.282$ & $+1.178$ \\[1ex]
		\botrule
	\end{tabular*}
	\footnotetext{\textbf{Note}: The last column of the table display the $\Delta$DIC values relative to the $\Lambda$CDM model, derived from the same data combinations. $\chi_{md}^2$ represents the median value of $\chi^2$. The parameter $z_t^*$ is not a free parameter.}
	\label{tab-GRB-OHD-BAO}
\end{table*}

\subsection{Results from the simultaneous fitting method}
\begin{figure*}[htbp]
	\centering
	\includegraphics[width=0.49\textwidth]{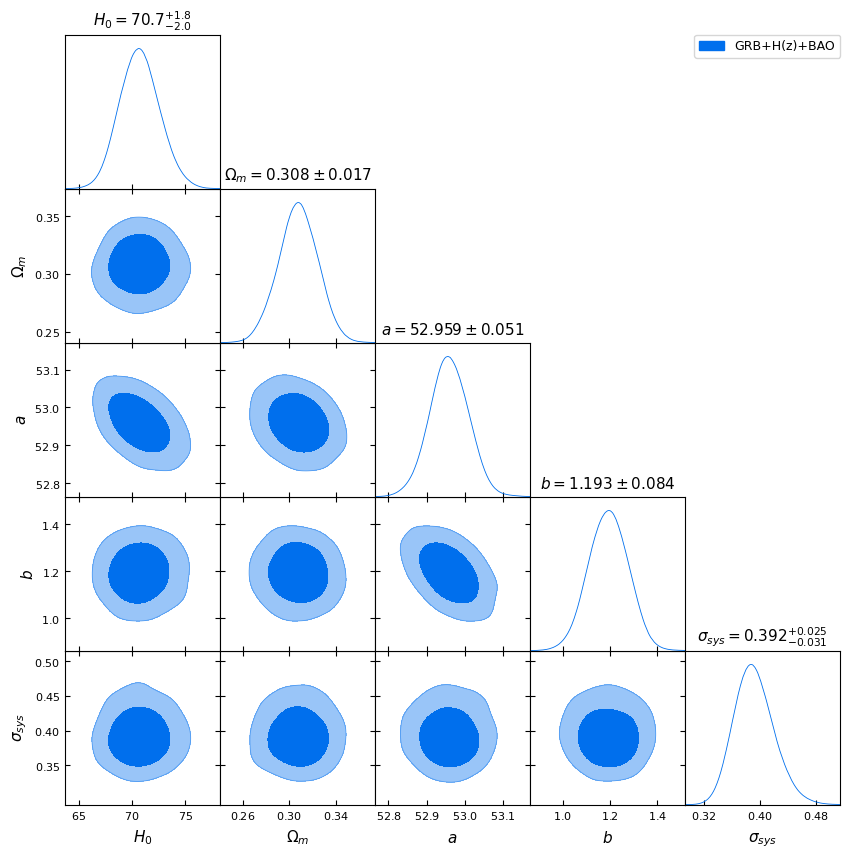}\\
	\includegraphics[width=0.49\textwidth]{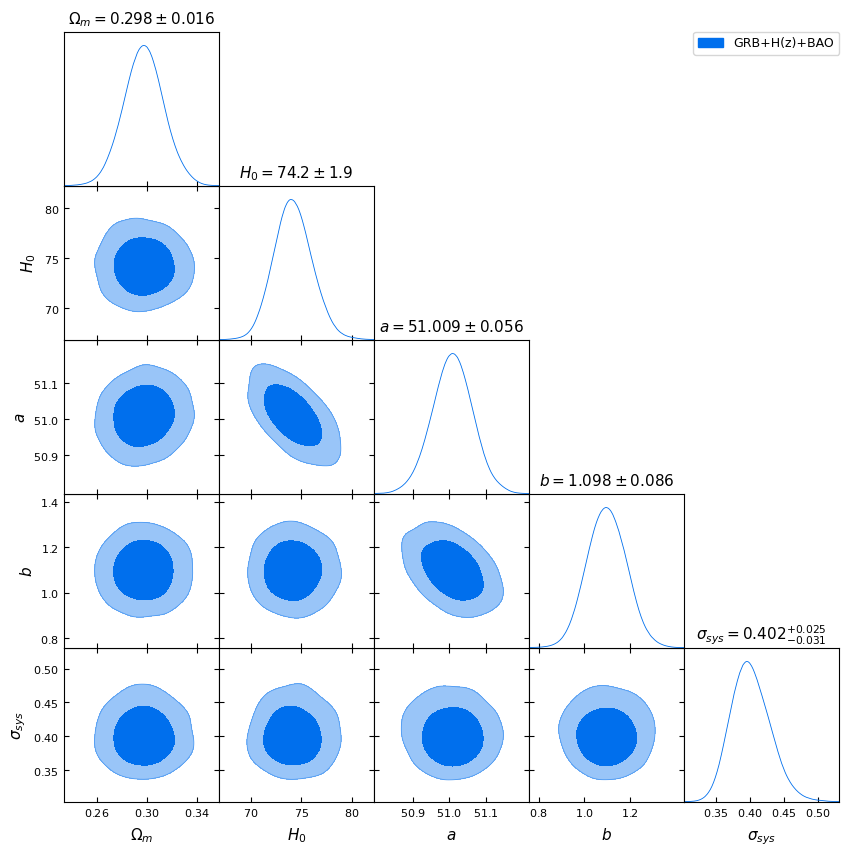}
	\includegraphics[width=0.49\textwidth]{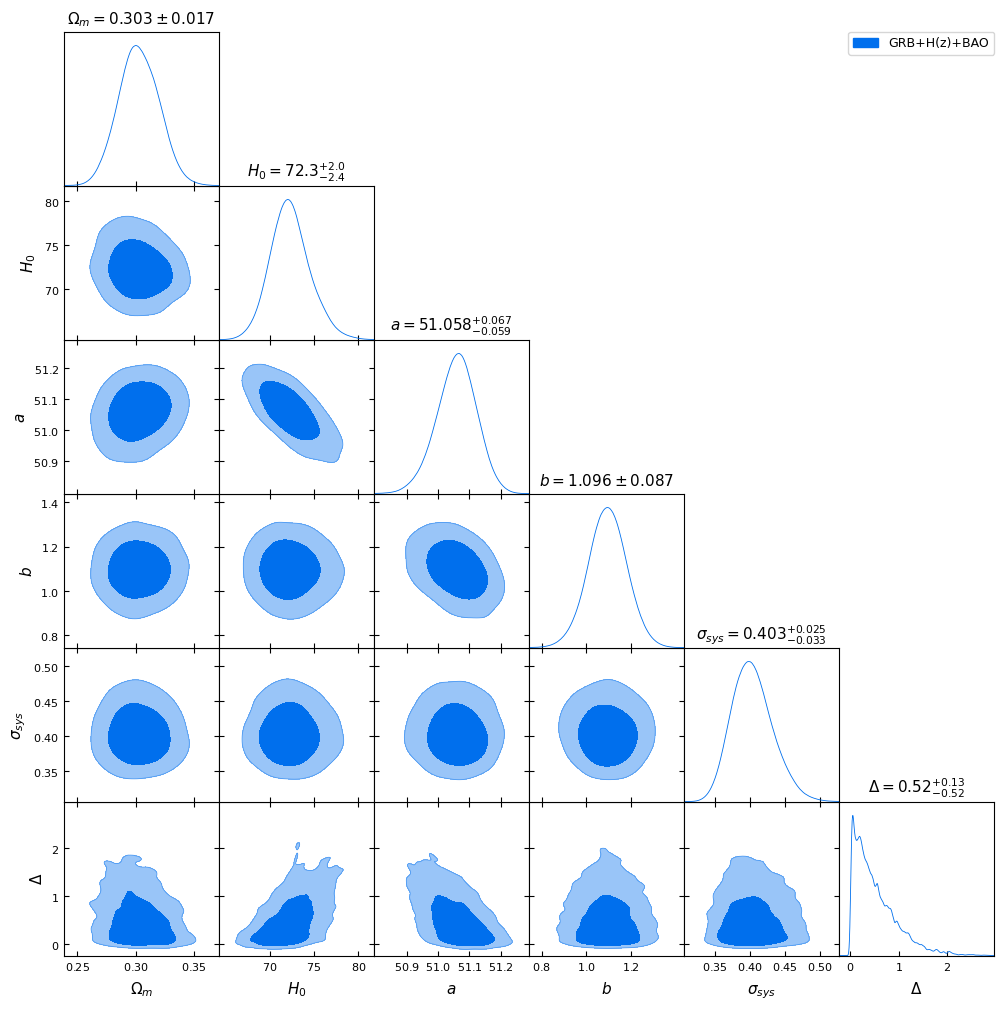}
	\caption{Simultaneous fitting of cosmological parameters ($\Omega_{m},H_0$) and GRB calibration parameters ($a,b,\sigma_{sys}$) with GRBs + OHD + BAO. The top panel is the $\Lambda CDM$, the bottom left is the PEDE, the bottom right is the GEDE model. \label{fig-sim}}
\end{figure*}

Finally, we constrain the $\Lambda$CDM, PEDE and GEDE models by using the simultaneous fitting method \citep{2019MNRAS.486L..46A}, in which  the parameters of DE models and the relation parameters of GRBs are fitted simultaneously.
Results with the data combination are shown in Fig. \ref{fig-sim}.  For the $\Lambda$CDM model, we obtain: 
$H_0=70.7_{-2.0}^{+1.8} \, \mathrm{km \, s^{-1} Mpc^{-1}}$ and $\Omega_{m}=0.308\pm0.017$, which are consistent with the results by \cite{2022ApJ...941...84L}. For PEDE model and GEDE model, $H_0=74.2 \pm 1.9 \, \mathrm{km \, s^{-1} Mpc^{-1}}$  for the PEDE model and  $H_0=72.3_{-2.4}^{+2.0} \, \mathrm{km \, s^{-1} Mpc^{-1}}$ for the GEDE model, respectively. Both EDE models yield similar results for the relation parameters, which are close to the local measurement with the SH0ES measurement.

\unskip

\section{Conclusions\label{section5}}

In this work, we have investigated the viability of the PEDE and the GEDE models with cosmology-independent observational data including GRBs and OHD samples.
The joint datasets of GRBs + OHD  in the mid-redshift region between the local the distance ladder SN Ia and CMB appear to provide much better constraints on the DE parameters and the value of $H_0$.
For comparison, we also consider the $\Lambda$CDM model, the $w$CDM model and the CPL  model.
With a Bayesian statistical approach for parameter inference and model selection,
we find that PEDE and GEDE  derive higher $H_0$ compare to $\Lambda$CDM, which support the viability of EDE models as a description of DE behavior and provide new evidence for their potential as an important supplement and possible alternative to the $\mathrm{\Lambda CDM}$ model.
Our results indicate that EDE models 
are at least competitive with the $\mathrm{\Lambda CDM}$ model in describing the accelerated expansion of the universe and can alleviate the $H_0$ tension problem, which are consistent with previous analyses \citep{Li_2019,Li_2020,2020JCAP...06..062P,2020MNRAS.497.1590H,2022A&A...668A..51L}.

In conclusion, our work demonstrates that the EDE models can better represent the effective behavior of DE compared to the $\mathrm{\Lambda CDM}$ model and can 
reduce tensions in the estimation of $H_0$.
Given the challenges faced by the standard cosmological model, this implies that EDE models can be competitive cosmological models.
Future theoretical explorations and observational verifications are needed to further test the validity of EDE models.
Recently, it should be note that the potential use of machine learning (ML) algorithms for cosmological use with GRBs \citep{2021MNRAS.503.4581L,2023arXiv231209440Z,2024arXiv240117029S} and OHD \citep{2023EPJC...83..548B,2023PhRvD.108j3526G}.
Moreover, the running of $H_0$  evolving with redshift is an interesting idea for the $H_0$ tension \citep{2021ApJ...912..150D,2022PhRvD.106d1301O,2023A&A...674A..45J,2023Univ....9...94H}.  \cite{2024A&A...681A..88H}  tested the cosmological principle by the region fitting (RF) method with Pantheon+ sample to simultaneously map matter-density distribution and the Hubble expansion distribution. The results provide clear
indications for a possible cosmic anisotropy.
Only with support from multiple lines of evidence can we ultimately determine the status of EDE models in explaining the $H_0$ tension.

\section*{ACKNOWLEDGMENTS}
We thank HuiFeng Wang, Xiaodong Nong, Zhen Huang and Xin Luo for discussions.
We are also grateful to the reviewers for their comments and suggestions.

\noindent\textbf{Funding}  This project was supported by the Guizhou Provincail Science and Technology Foundation: QKHJC-ZK[2021] Key 020 and QKHJC-ZK[2024] general 443.
X. Li was supported by NSFC No. 12003006, Science Research Project of Hebei Education Department No. BJK2024134 and the fund of Hebei Normal University No. L2020B02.\\

\noindent\textbf{Data Availability} No datasets were generated or analysed during the current study.

\section*{Declarations}

\noindent\textbf{Competing interests} The authors declare no competing interests.\\

\noindent\textbf{Ethics approval} Not applicable.

\bibliography{refs}
\end{document}